\documentclass[conference]{IEEEtran}
\IEEEoverridecommandlockouts
\usepackage{cite}

\usepackage{amsmath,amssymb,amsfonts}
\usepackage{algorithmic}
\usepackage[nolist]{acronym}
\usepackage{graphicx}
\usepackage{textcomp}
\usepackage{xcolor}
\usepackage{url}
\usepackage[colorinlistoftodos]{todonotes}

\usepackage[utf8]{inputenc}
\usepackage{pgfplots}
\DeclareUnicodeCharacter{2212}{−}
\usepgfplotslibrary{groupplots,dateplot}
\usetikzlibrary{patterns,shapes.arrows}
\pgfplotsset{compat=newest}

\usepackage{bm,amsmath,amssymb}
\newcommand{\task}{b}

\newcommand{\VNR}{v}
\newcommand{\VNRlist}{V}
\newcommand{\VNRduration}{\delta}
\newcommand{\minDuration}{\delta_\mathrm{min}}
\newcommand{\maxDuration}{\delta_\mathrm{max}}

\newcommand{\VNRpriority}{\lambda}
\newcommand{\minPriority}{\lambda_\mathrm{min}}
\newcommand{\maxPriority}{\lambda_\mathrm{max}}

\newcommand{\taskOne}{\mathrm{K}}
\newcommand{\taskTwo}{\mathrm{L}}
\newcommand{\taskThree}{\mathrm{M}}
\newcommand{\tasks}{B}
\newcommand{\capacity}{c_\mathrm{req}}

\newcommand{\link}{e}
\newcommand{\links}{E}

\newcommand{\linkOne}{\mathrm{KL}}
\newcommand{\linkTwo}{\mathrm{LM}}
\newcommand{\rateReq}{r_\mathrm{req}}

\newcommand{\node}{p}
\newcommand{\nodeSet}{P}

\newcommand{\nodeCapacity}{C}
\newcommand{\nodePower}{S}
\newcommand{\nodeNoise}{N_0}

\newcommand{\interference}{I}
\newcommand{\attenuation}{\gamma}

\newcommand{\rateMax}{r_\mathrm{max}}
\newcommand{\bandwidth}{\mathrm{BW}}
\newcommand{\timeslots}{T}

\newcommand{\placeVar}{\theta}
\newcommand{\routeVar}{\eta}

\newcommand{\extrareward}{r_\mathrm{extra}}
\newcommand{\weightDuration}{f_\VNRduration}
\newcommand{\weightPriority}{f_\VNRpriority}
\newcommand{\constantDuration}{c_\VNRduration}
\newcommand{\constantPriority}{c_\VNRpriority}
\usepackage{siunitx}  

\usepackage[caption=false,font=footnotesize]{subfig}

\newcommand\copyrighttext{%
  \footnotesize  © This work has been accepted in IEEE ANTS 2021.  Personal use of this material is permitted.  Permission from IEEE must be obtained for all other uses, in any current or future media, including reprinting/republishing this material for advertising or promotional purposes, creating new collective works, for resale or redistribution to servers or lists, or reuse of any copyrighted component of this work in other works}
\newcommand\copyrightnotice{%
\begin{tikzpicture}[remember picture,overlay]
\node[anchor=south,yshift=10pt] at (current page.south) {\fbox{\parbox{\dimexpr\textwidth-\fboxsep-\fboxrule\relax}{\copyrighttext}}};
\end{tikzpicture}%
}

\def\BibTeX{{\rm B\kern-.05em{\sc i\kern-.025em b}\kern-.08em
    T\kern-.1667em\lower.7ex\hbox{E}\kern-.125emX}}
\begin{document}

\title{Reinforcement Learning for Admission Control in Wireless Virtual Network Embedding}

\author{
   \IEEEauthorblockN{Haitham Afifi}
   \IEEEauthorblockA{
     Hasso Platter Institute \& \\
     Potsdam  University\\
             Potsdam, Germany \\
             haitham.afifi@hpi.de
           }
   \and
   \IEEEauthorblockN{Fabian Sauer}
   \IEEEauthorblockA{ 
                   Paderborn University\\
                     Paderborn, Germany \\
                    fjsauer@mail.uni-paderborn.de}
   \and 

   \IEEEauthorblockN{Holger Karl}
   \IEEEauthorblockA{  
     Hasso Platter Institute \& \\
     Potsdam  University\\
             Potsdam, Germany \\
             holger.karl@hpi.de
                             }
}

\maketitle
\copyrightnotice

\begin{acronym}
    \acro{RL}{Reinforcement Learning}
    \acro{NFV}{Network Function Virtualization}
    \acro{SFC}{Service Function Chain}
    \acro{VNR}{Virtual Network Request}
    \acro{VNE}{Virtual Network Embedding}
    \acro{WSN}{Wireless Sensor Network}
    \acro{TDMA}{Time Division Multiple Access}
    \acro{MAC}{Medium Access Control}
    \acro{IoT}{Internet of Things}
    \acro{2D}{two-dimensional}
    \acro{ID}{Identifier}
    \acro{PPO1}{Proximal Policy Optimization 1}
    \acro{PPO}{Proximal Policy Optimization}
    \acro{A2C}{Advantage Actor Critic}
    \acro{ACER}{Actor-Critic with Experience Replay}
    \acro{ACKTR}{Actor Critic using Kronecker-Factored Trust Region}
    \acro{DDPG}{Deep Deterministic Policy Gradient}
    \acro{DQN}{Deep Q Network}
    \acro{GAIL}{Generative Adversarial Imitation Learning}
    \acro{TRPO}{Trust Region Policy Optimization}
    \acro{SAC}{Soft Actor Critic}
    \acro{TD3}{Twin Delayed DDPG}    
    \acro{QoS}{Quality of Service}    
\end{acronym}

\begin{abstract}

Using Service Function Chaining (SFC) in wireless networks became popular in many domains like networking and multimedia. It relies on allocating network resources to incoming SFCs requests, via a Virtual Network Embedding (VNE) algorithm, so that it optimizes the performance of the SFC. When the load of incoming requests -- competing for the limited network resources -- increases, it becomes challenging to decide which requests should be admitted and which one should be rejected.

In this work, we propose a deep Reinforcement learning (RL) solution that can learn the admission policy for different dependencies, such as the service lifetime and the priority of incoming requests. We compare the deep RL solution to a first-come-first-serve baseline that admits a request whenever there are available resources. We show that deep RL outperforms the baseline and provides higher acceptance rate with low rejections even when there are enough resources. 

\end{abstract}

\begin{IEEEkeywords}
    acceptance rate maximization, wireless sensor network, reinforcement learning
\end{IEEEkeywords}
\section{Introduction}

\ac{NFV} is a common trend in networking as seen in telecommunications~\cite{sfc_survey}, Multimedia~\cite{streaming_sfc} and cloud computing~\cite{MlForWsnSurvey}. One of its key roles is to embed a chain of services, also known as \ac{SFC}, into infrastructure networks. The process of embedding is defined as \ac{VNE}, where a controller receives a \ac{VNR} to allocate resources (e.g., CPU capacities and routes) to these services.

It becomes challenging when these \acp{VNR} have different minimum resource requirements. In this case, multiple \acp{VNR} are competing for the finite resources of infrastructure networks. A \ac{VNE} algorithm can only allocate resources to optimize the performance of new incoming \acp{VNR}, but it cannot terminate an embedded \ac{VNR} to accept new ones. Consequently, we need to control the admission of these \acp{VNR} into the infrastructure to decide which \acp{VNR} should be embedded. A possible naive, greedy, solution is to admit a \ac{VNR}  that can be served at the time of request, i.e., First Come First Serve. 

Meanwhile, when the services differ in required resources, value or importance, the naive solution runs the risk of admitting less valuable, more resource-hungry services only to reject several other upcoming \acp{VNR}. We are hence interested in an admission control approach for \acp{VNR}, which optimizes the long-term average of admitted services (possibly weighted by revenue, value, priority, etc.) rather than just myopically focusing on the current request. 

As this needs an understanding of upcoming requests, explicit models for that are usually not available, reinforcement learning is a promising approach.	In this work, we use \ac{VNE} in \ac{WSN} as a case study and use two parameters to describe incoming \acp{VNR}:
\begin{itemize}
	\item service lifetime: duration of leasing the network resources
	\item priority of embedding a \ac{VNR} 
\end{itemize} 

Such applications are seen in smart environments with few to many distributed smart devices. These devices are supported by sensors that generate \acp{VNR} for heavy processing (seen in, e.g., gaming and multimedia applications) that can be executed on the near-by smart devices.  
We hence investigate RL as a tool for admission control of \acp{VNR}, where the uncertainty of upcoming \acp{VNR} is the main challenge. This could be related to, for example, changing arrival rates or  \acp{VNR} parameters. We focus here on the latter, so that  \acp{VNR} have different service lifetime and different priorities. The main objective of  admission control is then to embed as many \acp{VNR} as possible (or  \acp{VNR} with higher revenues), averaged over long time. For fairness and to avoid abusing the resources, an embedded \ac{VNR} is terminated if its service lifetime ends.


	In the following sections, we give a brief overview of related work with respect to admission control and \ac{RL}, and how it differs from our work (Section~\ref{sec:related_work}). Then, we formulate the  problem  and summarize the applied \ac{VNE} solution (Section~\ref{sec:problem_formulation}). Next, we describe our \ac{RL} framework (Section~\ref{sec:reinforcement_learning}) and compare the proposed solution to greedy admission control in  different simulation setups (Section~\ref{sec:simulation_setup}). Finally, we summarize the outcome in Section~\ref{sec:summary}


\section{Related work}
\label{sec:related_work}

Using an \ac{RL} approach in the context of \acp{WSN} has been used to solve many problems as in \ac{MAC}~\cite{CrWsnRl}, energy saving~\cite{CoverageWsnRl}, and many other similar problems~\cite{MlForWsnSurvey}. Meanwhile, we focus here on work that explicitly consider \ac{RL}, \ac{VNE} and admission control.


Different types of requests were investigated when using \ac{RL} for admission control. For instance, the authors in~\cite{asif_ac_RL,rl_flowcontrol,rl_completitionTime} assume that incoming requests have flow properties that need to be completed before some deadline, while the authors in~\cite{rl_job_ac,rl_job_ac2} assume that incoming requests are jobs that need to be running on servers, e.g., cloud or edge servers. In our work, the requests have the properties of both jobs and flows, since a \ac{VNR} has multiple jobs connected via link flows.

The authors in \cite{early_rl_ac,ac_rl_qos} assume that  \ac{QoS} is a constraint, hence, their proposed solutions reject \acp{VNR} that are likely to violate \ac{QoS} bounds. In contrast to their assumption, we assume that \ac{QoS} is an objective for the \ac{VNE} problem, while the admission control maximizes the embedding revenue.

 Furthermore, the work in~\cite{ac_rl_qos} assumed arriving \acp{VNR} wait in a queue for a decision to be embedded or to be rejected. Meanwhile, in our work, and similar to~\cite{early_rl_ac},  we have a queue length of only one. Therefore, queuing issues have been ignored to emphasize other features of the admission control problem. Consequently, using \ac{RL} for optimizing waiting time in the queue~\cite{rl_completitionTime,rl_waitingTime}, queue length~\cite{dql_comm} and queuing management~\cite{rl_traffic_engine} is beyond the scope of this paper.

 
 We use a \ac{VNE} solution to check the feasibility of embedding a \ac{VNR}. To combine both admission control and \ac{VNE} problems, the work in~\cite{ac_vne_rnn} uses recurrent neural networks to reject \acp{VNR} that are likely to fail the \ac{QoS} constraints, which saves the computational time needed by the \ac{VNE} solution to check the \ac{VNR} implementation feasibility. Hence, the admission control was trained to have high accuracy of detecting feasible accepted \ac{VNR} embedding. In our work, the admission control is trained to maximize the network's revenue, meaning that it can reject feasible \acp{VNR} in order to accept more \acp{VNR} later. Using \cite{ac_vne_rnn} as a quick feasibility check and then using our RL agent to reject unpromising requests should make an interesting followup study. 
 
 Another prediction model was used by~\cite{rl_ac_arrivalRate} to predict the arrival rate of upcoming \acp{VNR}. This is then followed by an \ac{RL} agent, whose objective is to maximize the acceptance rate. In contrast, we do not use the arrival rate as an observation so that it is being implicitly learned by the \ac{RL} agent. Building up on our work to consider changing arrival rates is straight forward, yet it requires deep analyses with respect to the uncertainty of the changes and the performance, which we leave as an extension to this work.
 
 Further combinations of admission control and \ac{VNE} decisions have been solved using \ac{RL} in~\cite{ac_vne_rl,VnePreprocessorProposal}. Similar to our work, the objective is to maximize the network's revenue. The monolithic mix of \ac{VNE} and admission control applied in~\cite{ac_vne_rl} is valid only for wired networks and cannot be applied or reused in our wireless network, due to the differences between wired and wireless \ac{VNE} problems~\cite{Marvelo}. However, we assume a modular implementation, where admission control and \ac{VNE} solutions are two separate modules that interact with each other. This allows reusing different modules in similar problems. Additionally, the simplicity of our problem will probably lead to  shorter training time and faster convergence~\cite{large_action_space_conv}.

\section{Problem formulation}
    \label{sec:problem_formulation}
    
\begin{figure}
    	\centering
    	\includegraphics[clip, trim=2cm 3cm 2cm 2cm,width=0.46\textwidth]{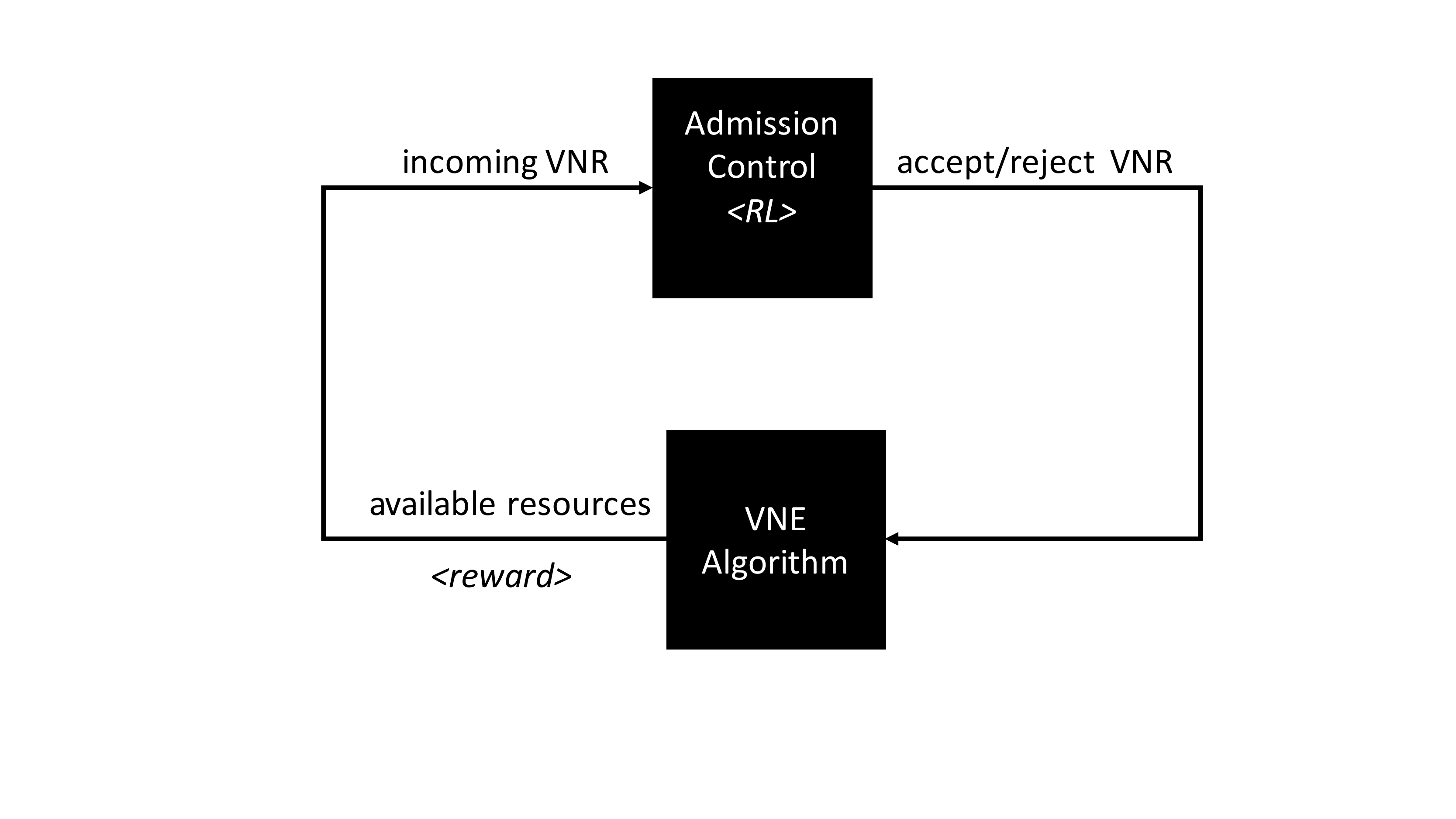}
    	\caption{Interaction between VNE and admission control}
    	\label{fig:vne-ac}
    \end{figure}

    There are two objectives for our problem. First, we need to maximize the revenue from accepting \acp{VNR}, which is based on the acceptance rate, service time or/and priority of \acp{VNR}. Second, we need to minimize the number of used time slots per an accepted \ac{VNR} for a high \ac{QoS}. The latter is solved by a heuristic \ac{VNE} solution (Section~\ref{sec:vne_solution}). Because of our modular implementation, this \ac{VNE} solution is treated as a black box to our first objective (Figure~\ref{fig:vne-ac}).

    The first objective --our main focus in this paper-- would require rejecting \acp{VNR}, even if they could be embedded by the \ac{VNE} solution, to allow more/better upcoming \acp{VNR} to be embedded instead. Hence, the decisions of whether to accept/reject a \ac{VNE} depends on the resource allocation but not vice versa. 
    
    In the following subsections, we describe the features of the \acp{VNR} and the wireless network for formulating our problem, and describe briefly the \ac{VNE} solution, since it interacts with the admission control.



    \subsection{Virtual network requests}
        
        A \ac{VNR} consists of $\tasks$ tasks connected via $\links$ links. Each processing task $\task \in \tasks$ requires a capacity $\capacity^\task$, while 
        a link has a
        minimum transmission data rate $\rateReq$ requirement.

    We assume that we have a predefined set of \acp{VNR}, whose typologies are known in advance. Hence, each \ac{VNR} is labeled with an \ac{ID}, where the duration and priority of each \ac{VNR} can change, but the topology and required resources do not.
    
    Additionally, we assume that we have a discrete time environment, where at each time slot a new \ac{VNR} $\VNR$ arrives, whose runtime
     $\VNRduration^\VNR$ is uniformly distributed between $\left[\minDuration, \maxDuration\right]$ time steps.
    By altering the value of $\maxDuration$, the average \ac{VNR} duration can be adjusted to simulate different loads. Similarly, each \ac{VNR} has  a priority $\VNRpriority^\VNR \in \left[\minPriority,\maxPriority\right]$.

%
    
    \subsection{Wireless Sensor Network}
        The wireless sensor nodes operate in half-duplex communication mode; a node can be in one of three states: send, receive, idle.
        Each node $\node \in \nodeSet$ is defined using this set of properties:
        position in network computational capacity $\nodeCapacity^\node$, transmit power $\nodePower^\node$ and noise floor $\nodeNoise$.

        We assume that all sensor nodes share the same collision domain with bandwidth~$\bandwidth$. To allow multiple channel access, we assume an ideal TDMA with no collision and all transmissions are synchronized, in which we have $\timeslots$ time slots.

		Given the attenuation  $\attenuation^{i,j}$ between two nodes $i,j \in \nodeSet$ , the maximum achievable data rate $\rateMax^{i,j,t}$  at time slot $t\in \timeslots$ is given by

        \begin{equation}
            \rateMax^{i,j,t} = \frac{\bandwidth}{|\timeslots|} \mathrm{log_2}(1 + \frac{\nodePower^i\attenuation^{i,j}}{\interference^{i,j,t} + \nodeNoise})
            \label{eq:r_max}
        \end{equation}
    \noindent where $\interference^{i,j,t} = \sum_{\substack{\node\in \nodeSet^t \\ \node \neq i}} \attenuation^{\node,j}\nodePower^\node$ is the interference at node $j$ from other nodes $\nodeSet^t$ that are simultaneously transmitting with node $i$ at time slot $t$.

\subsection{Constraints}
		The required constraints for successful wireless \ac{VNE} are described in details in~\cite{Marvelo}, but we summarize them here whilst extending the formulation to a discrete time horizon.

		First, a successfully embedded \ac{VNR} $\VNR$ will decrease its time duration $\VNRduration^\VNR = \VNRduration^\VNR -1$
		each time step, so that $\VNR$ is running as long as $\VNRduration^\VNR >0$, otherwise, the \ac{VNR} will be terminated to avoid additional costs or abusing the resources. 
		Second, let us assume that we have a list of \acp{VNR} $\VNRlist^+$ which are currently running inside the network (i.e., $\VNRduration^\VNR >0, \quad \forall \VNR\in \VNRlist^+$). Then, we need to ensure the following

		\footnotesize
		\begin{align}
			\sum_{\nodeSet}\placeVar(\task^\VNR,\node)&=&1&,\quad \forall \task\in \VNR, \forall \VNR \in \VNRlist^+ \label{eq:placed} \\
			\sum_{\task \in  \tasks }\placeVar(\task^\VNR,\node)\capacity^\task&\leq&\nodeCapacity^\node&,\quad  \forall \VNR \in \VNRlist^+,\forall \node \in \nodeSet \label{eq:capacity} \\
			\routeVar(\link^\VNR,i,j,t)\rateReq^{\link^\VNR}&\leq&\rateMax^{i,j,t},& \quad \forall \left\{i,j\right\}\in \nodeSet ,t\in\timeslots, \link \in \links, \VNR \in \VNRlist^+~\label{eq:rateLimit}
		\end{align}
		\normalsize
		
		We use two binary variables to formulate the constraints: $\placeVar(\task,\node)$ and $\routeVar(\link ,i,j,t)$. The former is used to state if task $\task$ is running on node $\node$, while the latter is used to state if node $i$ is transmitting to node $j$ the data of link $\link$ at time slot $t$.
		
		  In Eq.~\eqref{eq:placed}, we ensure that each task $\task \in \tasks$ where $\VNR \in \VNRlist^+$ is running on a node. However, all nodes should not be overutilized (Eq.~\eqref{eq:capacity}). Similarly, to guarantee an upper-bound delay, Eq.~\eqref{eq:rateLimit} checks, if node $i$ is transmitting the link $e$ to node $j$ at time slot $t$, then $\rateReq^e$ is less than or equal to the maximum achievable rate $\rateMax^{i,j,t}$. Note that we assume, for simplicity, a perfect medium access channel with constant channel access delay. Based on the interference from different transmissions, two nodes may transmit data of different links simultaneously.
		  
		In addition to the above constraints, we check the flow conservation ones to ensure successful routing. Meanwhile, we drop the formulation of these constraints lest we distract from the goal of the paper; the reader is referred to \cite{wmnc_2019} for a detailed description.

\subsection{\ac{VNE} solution}
\label{sec:vne_solution}

Our \ac{VNE} solution is a straightforward first-fit constructive heuristic~\cite{wmnc_2019}, which finds a solution by following a sequence of pre-ordered constraints. We summarize the process as following. First, wireless nodes are chosen at random for running the tasks on nodes, while checking the capacity constraint. If any node does not satisfy the capacity constraint, another node is selected at random. Next, we find shortest path routes between the nodes running the tasks. At the end, time slots are allocated to the transmissions between the nodes in a topological order. We start at the beginning with one time slot. If simultaneous transmissions cannot take place within the available time slots --i.e., due to duplex or $\mathrm{SINR}$ constraints --new additional time slots are used for transmissions. The optimality gap of this heuristic has been derived in~\cite{wmnc_2019}.

\section{Reinforcement learning for admission control}
\label{sec:reinforcement_learning}

In this section, we describe the \ac{RL} implementation of the admission control. As shown in Figure~\ref{fig:vne-ac}, the \ac{VNE} solution, described in Section~\ref{sec:vne_solution}, acts as an \ac{RL} environment. Accordingly, it additionally provides the controller with a reward for training purposes.
    
    In the following subsections, we define the observation space, action space and  reward function of the \ac{RL} environment.

\subsection{Observation space}
All observations are stored in a fixed-size multi discrete vector containing the following information: 

\begin{itemize}
	\item Node capacities: $\rightarrow \mathbb{R}^\nodeSet$
	\item Edge activation $\equiv \routeVar$ $\rightarrow$ $\mathbb{R}^{\nodeSet\times \nodeSet \times T \times 2}$
	\item New \ac{VNR} properties $\rightarrow \mathbb{R}^2$
	\begin{itemize}
		\item Service time
		\item Priority
	\end{itemize}
\end{itemize}

Node capacities contain the  available capacities of each node at the current time step.
Edge activation is a multi-dimensional matrix containing all currently embedded links. 
The first and second dimension encode the edge of sending and receiving nodes.
The third dimension stands for the time step at which the transmission takes place.  Each transmission is labeled by 2 \acp{ID} representing which \ac{VNR} and which link within the \ac{VNR}. For new incoming \acp{VNR}, we add the \ac{VNR}~\ac{ID}, duration $\VNRduration^\VNR$ and priority $\VNRpriority^\VNR$

    \subsection{Action space}
        The \ac{RL} agent
        decides whether an incoming \ac{VNR} is rejected or accepted (binary space). In case of acceptance, the \ac{VNE} algorithm will try to embed the \ac{VNR} into the network.

    \subsection{Reward function}
    \label{sec:reward_function}
        The reward function is modeled to train the agent towards the desired behavior described in \ref{sec:problem_formulation}.
        \begin{table}[!htbp]
            \caption{Reward function}
            \centering
            \begin{tabular}[h]{c|c|c|c}
                Agent decision & \ac{VNE} solution & Label & Reward\\
                \hline
                accepted & feasible & \textit{true positive} & +6\\
                accepted & infeasible & \textit{false positive} & -2\\
                rejected & feasible & \textit{false negative} & -1 + $\extrareward$\\
                rejected & infeasible & \textit{true negative} & \textpm 0\\
            \end{tabular}
            \label{tab:reward_labels}
        \end{table}

        Table \ref{tab:reward_labels} describes the
        combinations of agent decisions and \ac{VNE} algorithm solutions with their corresponding reward.
        An incoming \ac{VNR} is labeled as \textit{true positive}, if the agent decides to accept the\ac{VNR} and the \ac{VNE} algorithm successfully embedded it.
        The other  labels describe the remaining combinations of agent's decisions and the \ac{VNE} solution feasibility.
        Note that these labels are just used to highlight the differences in taken decisions between the \ac{RL} and first-come-first-serve solutions and do not reflect the performance of the \ac{RL} decisions.
        
        \textit{False negatives} receive a positive reward $\extrareward$ calculated using Eq.~(\ref{eq:extra_reward}), which stimulates accepting \acp{VNR} with high priority and low service lifetime.
        It relies on the relative delay ($\weightDuration^\VNR= \frac{\VNRduration^\VNR}{\maxDuration}$) and the relative priority ($\weightPriority^\VNR = \frac{\maxPriority- \VNRpriority^\VNR}{\maxPriority-1}$) of the rejected \ac{VNR} $\VNR$.
        \begin{equation}
            \extrareward^\VNR =\constantDuration \cdot \weightDuration^\VNR + \constantPriority \cdot\weightPriority^\VNR
            \label{eq:extra_reward}
        \end{equation} 		
        
        The control parameters $\constantDuration$ and $\constantPriority$ are used to tune the \textit{false negative} behavior on the extra reward. In other words, they tune the weights for acceptance rate and the revenue.

\section{Simulation setup}
\label{sec:simulation_setup}

\begin{figure}[htbp]
	\centering
	\includegraphics[clip, trim=0cm 6cm 0cm 5cm,width=0.46\textwidth]{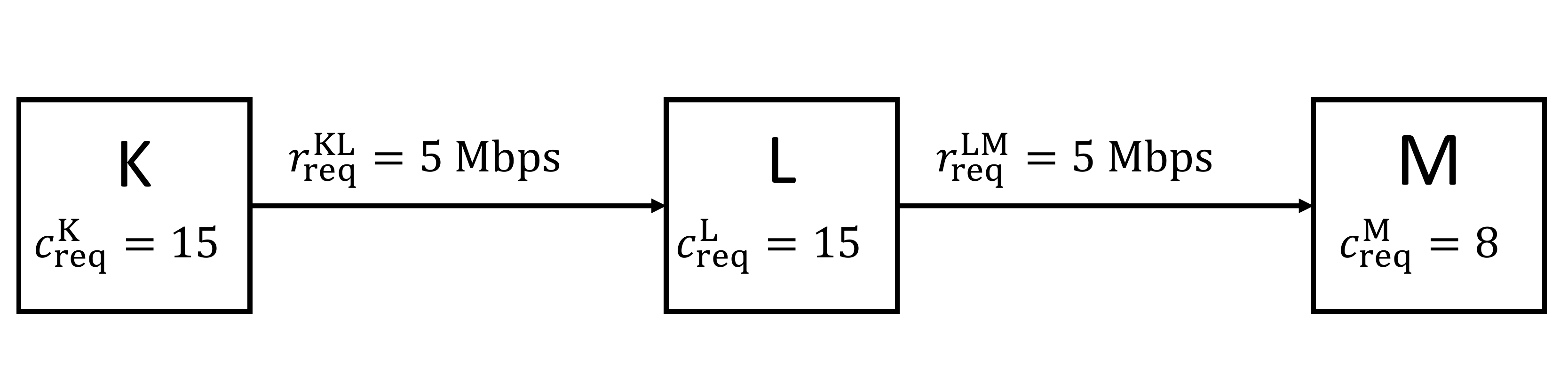}
	\caption{\ac{VNR} overlay graph}
	\label{fig:vnr}
\end{figure}

The \ac{VNR} used in this paper is made of three processing tasks  {${\tasks=\left\{\taskOne, \taskTwo, \taskThree \right\}}$}  connected via two links ($\links = \left\{ \linkOne, \linkTwo \right\}$) as shown in Fig. \ref{fig:vnr}.

    The number of time slots per time step is  $\timeslots = 8$ for all simulations in this paper.
    We simulate 5 nodes placed in a small room (e.g., an office) with dimensions $3\,\si{m}\,\times\,3\,\si{m},\times 3\,\si{m}$.
    The wireless channel bandwidth is set to $\bandwidth = 20\,\si{MHz}$. We set the minimum time duration per \ac{VNR} $\minDuration =2$ and $\maxDuration=30$, while the the \ac{VNR}'s priority $\VNRpriority \in \left[1,10\right]$.

    All trained agents are compared to a baseline that \textit{always accepts} a \ac{VNR} and embed it whenever there are enough resources. Accordingly, this will have the first-come-first-serve behavior, where the embedding of a \ac{VNR} depends only on the \ac{VNE} solution's feasibility. With respect to Table~\ref{tab:reward_labels}, the output of this algorithm corresponds only to \textit{true positives} and \textit{false positives}.

 To measure the impact of the control parameters ($\constantDuration$ and $\constantPriority$) and the sensitivity of the trained agent to incoming \ac{VNR} properties, we define 3 simulation setups:
 \begin{enumerate}
 	\item fix $\constantPriority$ and change $\constantDuration$
 	\item change $\constantPriority$ and fix $\constantDuration$
 	\item fix both $\constantPriority$ and $\constantDuration$, while evaluating different agents trained on different $\maxDuration$
 \end{enumerate}

\section{Simulation results}

	We have for each configuration setup 100 different runs, where each run has 1000 time steps (i.e., 1000 incoming \acp{VNR}).
	We compare the median --to exclude outliers-- of these runs to that from the baseline solution (gray dots).

    \subsection{Emphasizing low service time}
        \label{sec:cd_values}
\begin{figure}
   		\centering
   		\subfloat[Acceptance rate]{%
   			\centering
			\includegraphics[width=0.48\linewidth]{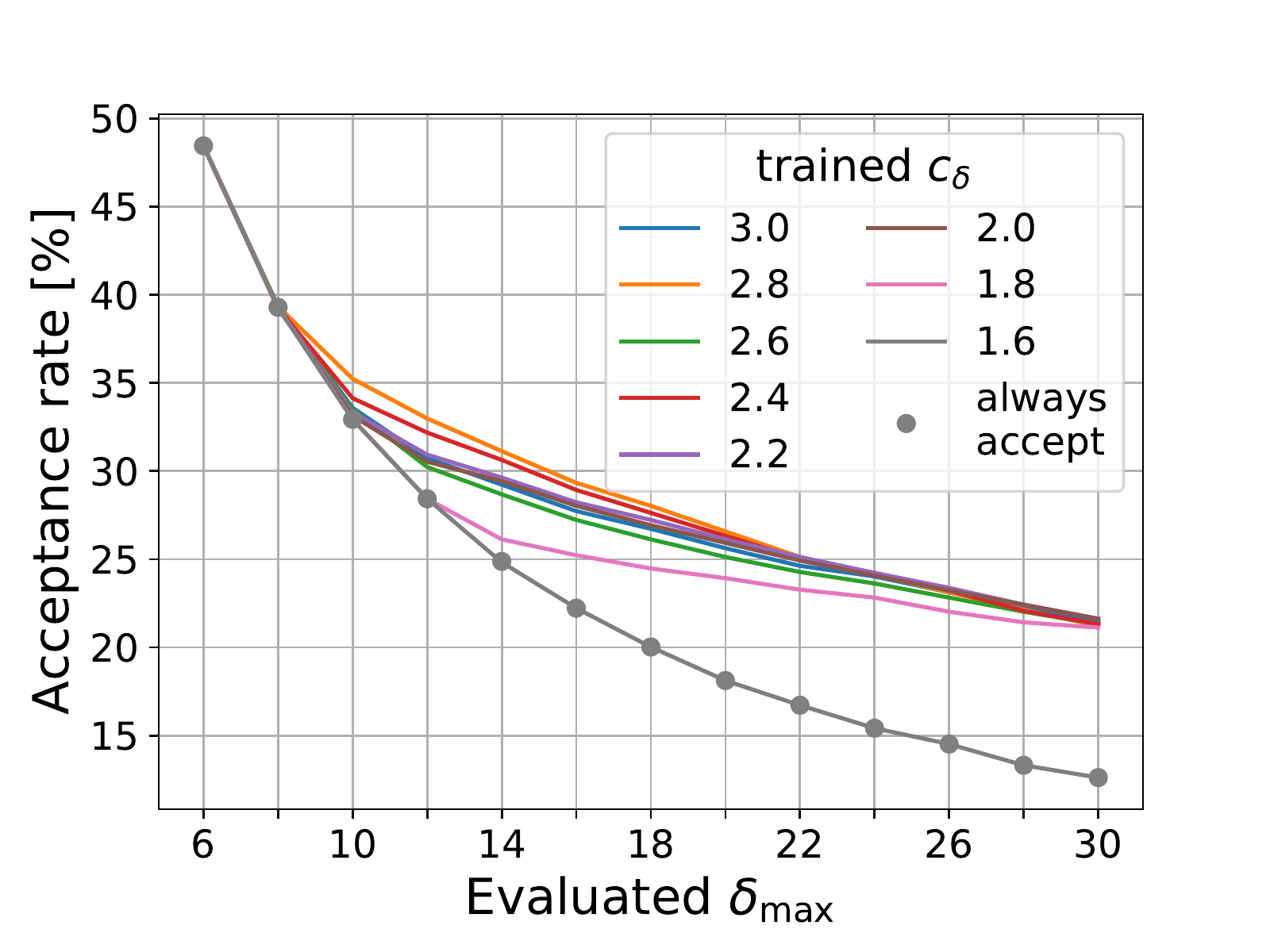}
   			\label{fig:tp_time}
   		}
   		\subfloat[False negatives]{%
   			\centering
   			\includegraphics[width=0.48\linewidth]{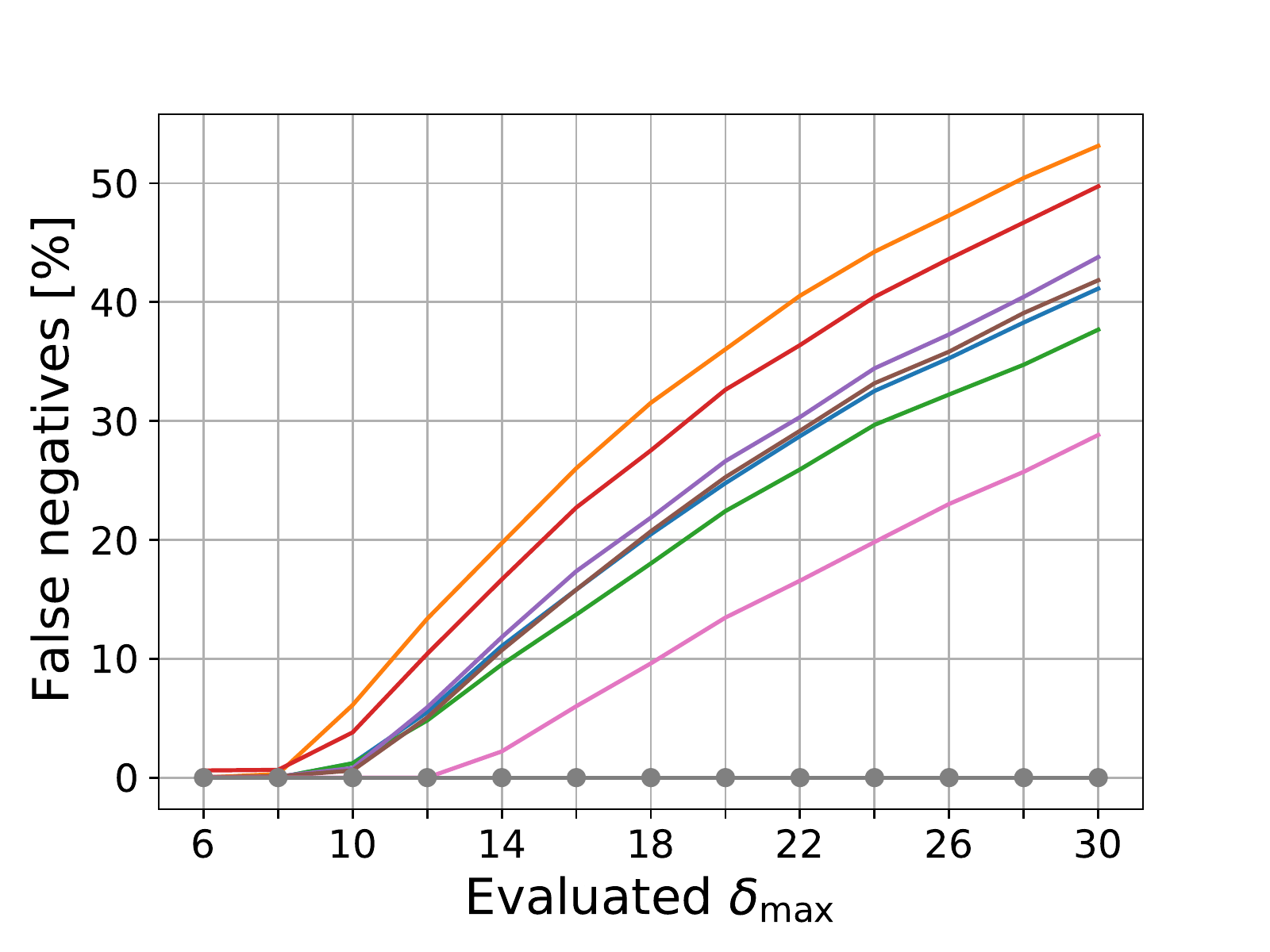}
   			\label{fig:fn_time}
   		}
   		\caption{Results of $\constantDuration$ parameter analysis}
   		\label{fig:c_time}
   	\end{figure}

        As stated in Section~\ref{sec:reward_function}, $\constantDuration$ controls the agent's decision (i.e., accept/reject a \ac{VNR} $\VNR$)  with respect to the \ac{VNR}'s duration $\VNRduration^\VNR$. To illustrate the sensitivity of agent's decision to this parameter, we train eight agents whose $\constantDuration$ change between $[1.6,3.0]$. All agents train for  the same number of steps $10^6$ and $\maxDuration=30$ (Fig.~\ref{fig:c_time}).
        
        
        Fig.~\ref{fig:tp_time} shows the median   acceptance rate ($\frac{\text{number of accepted \ac{VNR}}}{\text{total number of incomming \acp{VNR}}}$) of all trained agents and the \textit{always accept} baseline agent under different offered load.
        Fig.~\ref{fig:fn_time} shows the corresponding relative number of \textit{false negatives} created by those agents.
   	
        The agent trained with $\constantDuration = 1.6$ does not increase overall acceptance rate compared to the \textit{always accept} baseline agent and does not make any \textit{false negatives} decisions.
        Agents with higher values for $\constantDuration$ show an increase in overall acceptance rate especially in higher offered load environments.
        This increase comes with the cost of creating more \textit{false negatives}.
        The difference between each other is most present for medium offered loads.
        The agents with $\constantDuration= 1.8, 2.0, 2.2, 2.4$ show a gradual increase in acceptance rate compared to each other when evaluating $\maxDuration$ between 10 and 22. 
        For incoming \acp{VNR} with higher $\maxDuration$, the agents' perform closely to each other (as in  $c_\mathrm{d} = 2.6, 2.8, 3.0$) to be twice as high as the baseline, while the unnecessary rejections introduced by the \textit{false negatives} increases(Fig.~\ref{fig:fn_time}).
        
        Consequently,  $\constantDuration$ should be carefully tuned: very small values will have similar performance to the baseline solution (lower bound), while having high values will yield unnecessary rejections and minimal to no gain in the acceptance ratio.
        
    \subsection{Emphasizing high priorities}
\begin{figure}
            \centering
            \subfloat[Acceptance rate]{%
                \centering
            \includegraphics[width=0.48\linewidth]{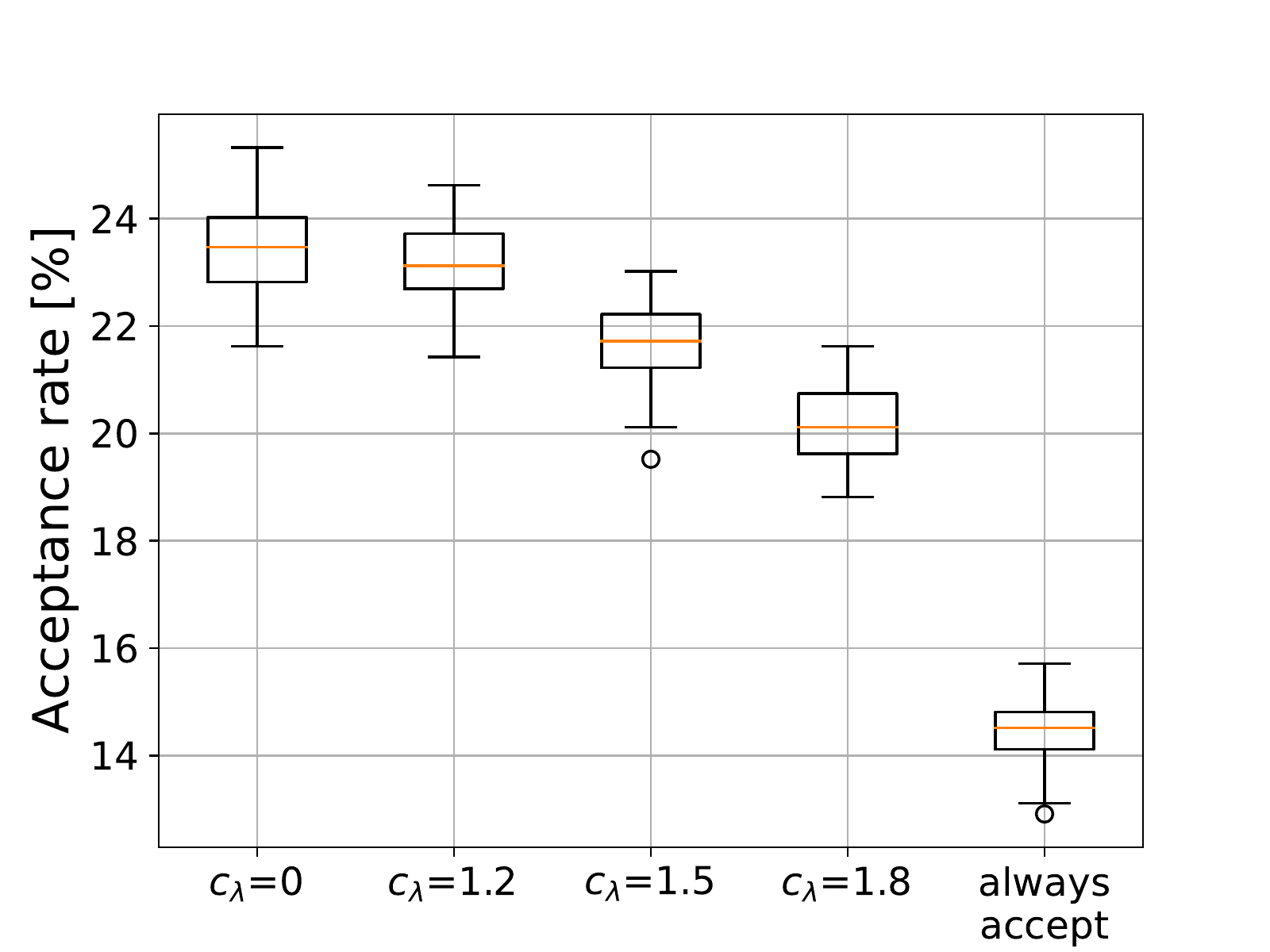}
                \label{fig:boxplot_tp}
            }
            \subfloat[False negatives]{%
                \centering
                \includegraphics[width=0.48\linewidth]{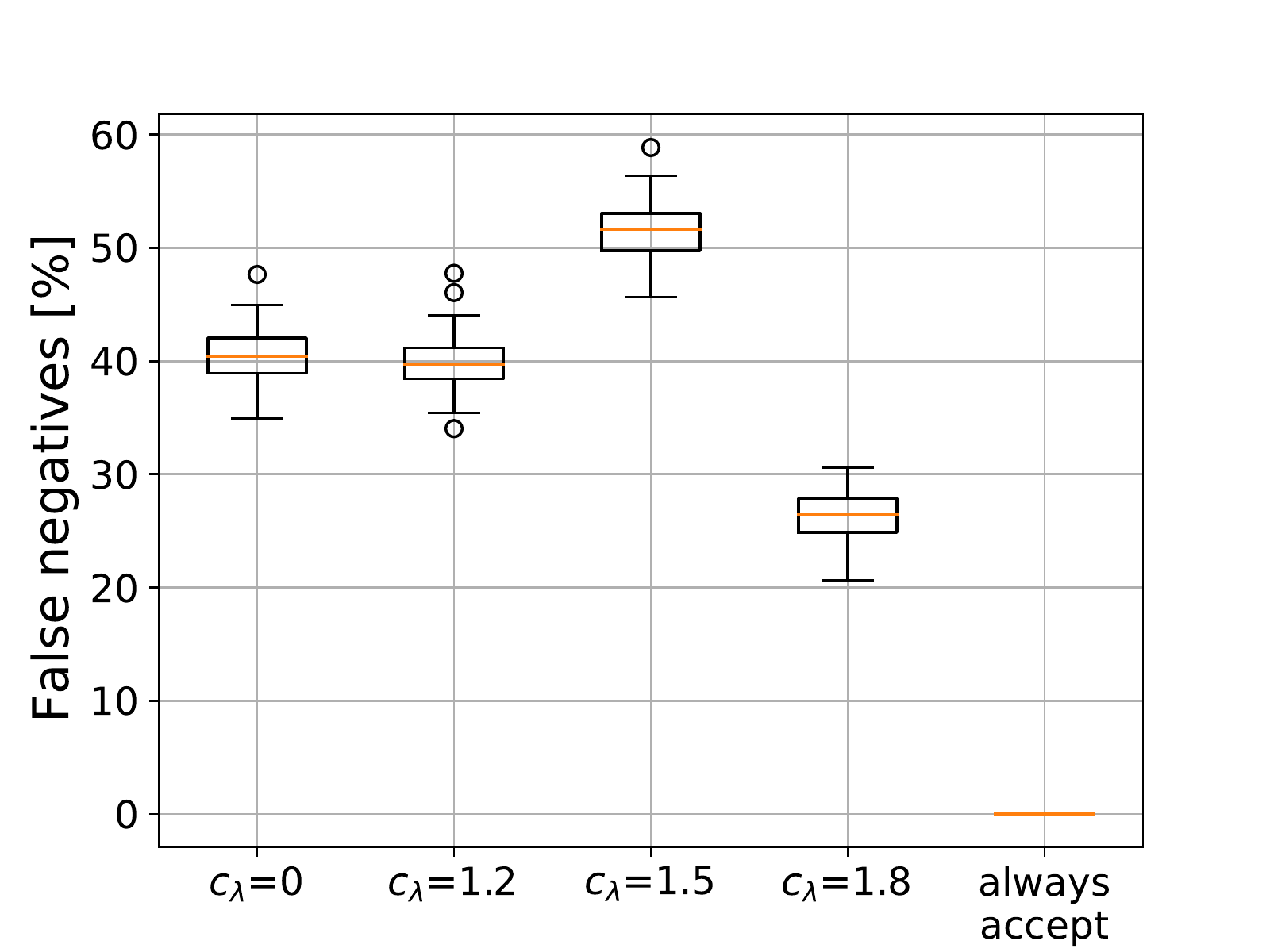}
                \label{fig:boxplot_fn}
            }
            \caption{Results of $\constantPriority$ parameter analysis with $\constantDuration = 1.8$ and $\maxDuration = 26$}
            \label{fig:boxplots}
        \end{figure}

        Previous analyses focused on maximizing the acceptance rate by rejecting long \acp{VNR}.
        This may not be ideal because it prevents longer \acp{VNR} from being embedded at all.
        To prevent this, $\constantPriority$ can be used to give higher priority for longer \acp{VNR} (e.g., higher revenue). 
        Fig.~\ref{fig:boxplots} shows the evaluation results of the trained agents with different values for $\constantPriority$ and $\VNRpriority \in [1,10]$, while all agents use $\maxDuration = 26$ for training and evaluations.
        
        In Fig.~\ref{fig:boxplot_tp}, we observe that the higher the value of $\constantPriority$ during training the lower the acceptance of the resulting agent. This is due to accepting \acp{VNR} with high $\maxDuration$ and high priority. Meanwhile, there is a tendency to decrease the number of \textit{false negatives} as $\constantPriority$ increases.
        
\begin{figure}[!t]
        	\centering
            \begin{tabular}[m]{cc}
                \begin{tabular}[m]{c}
                    \subfloat[True positives]{%
                        \centering
                        \includegraphics[width=0.45\linewidth]{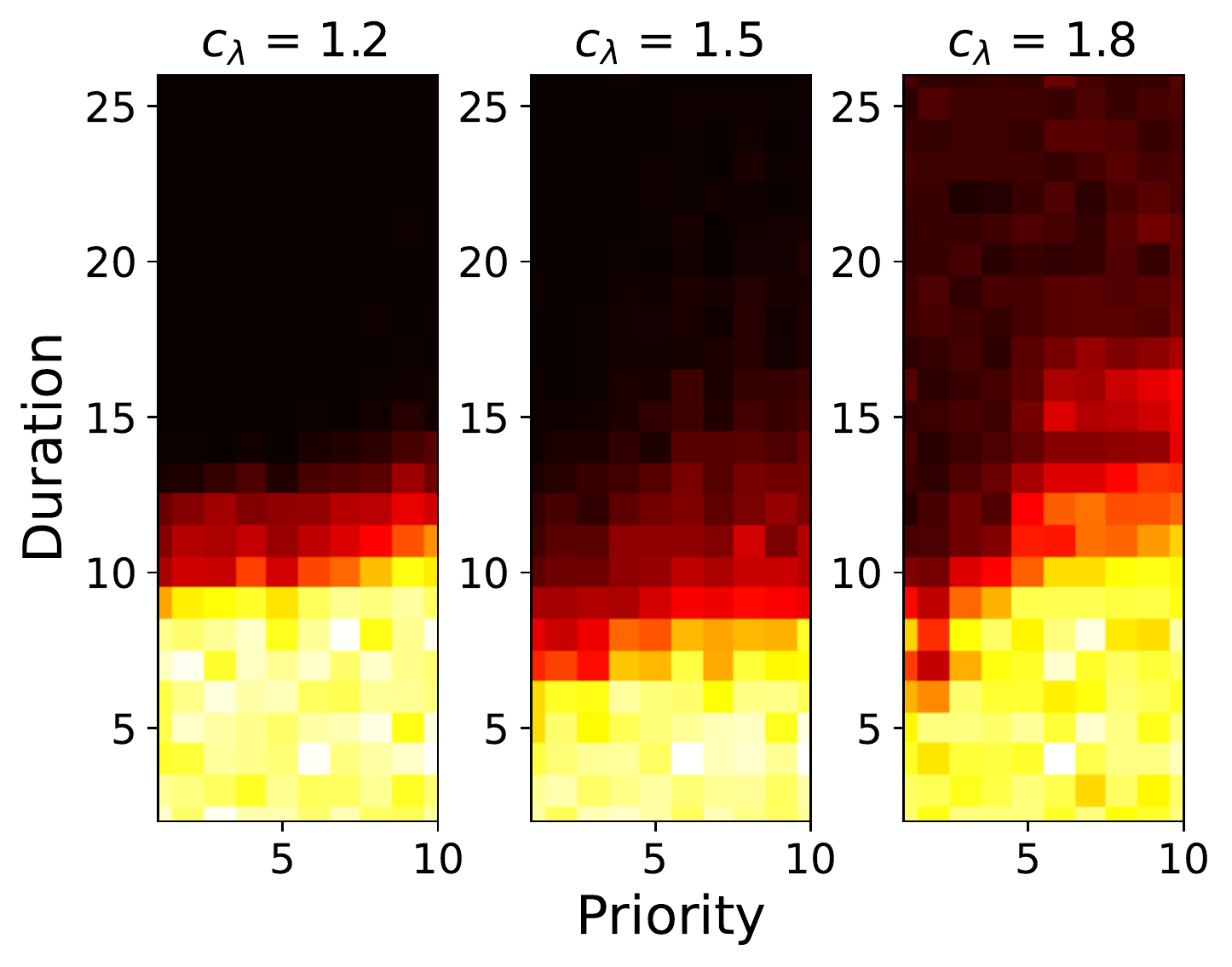}
                        \label{fig:cp_analysis_tp}
                    } \\
                    \\
                    \subfloat[False negatives]{%
                        \centering
                        \includegraphics[width=0.45\linewidth]{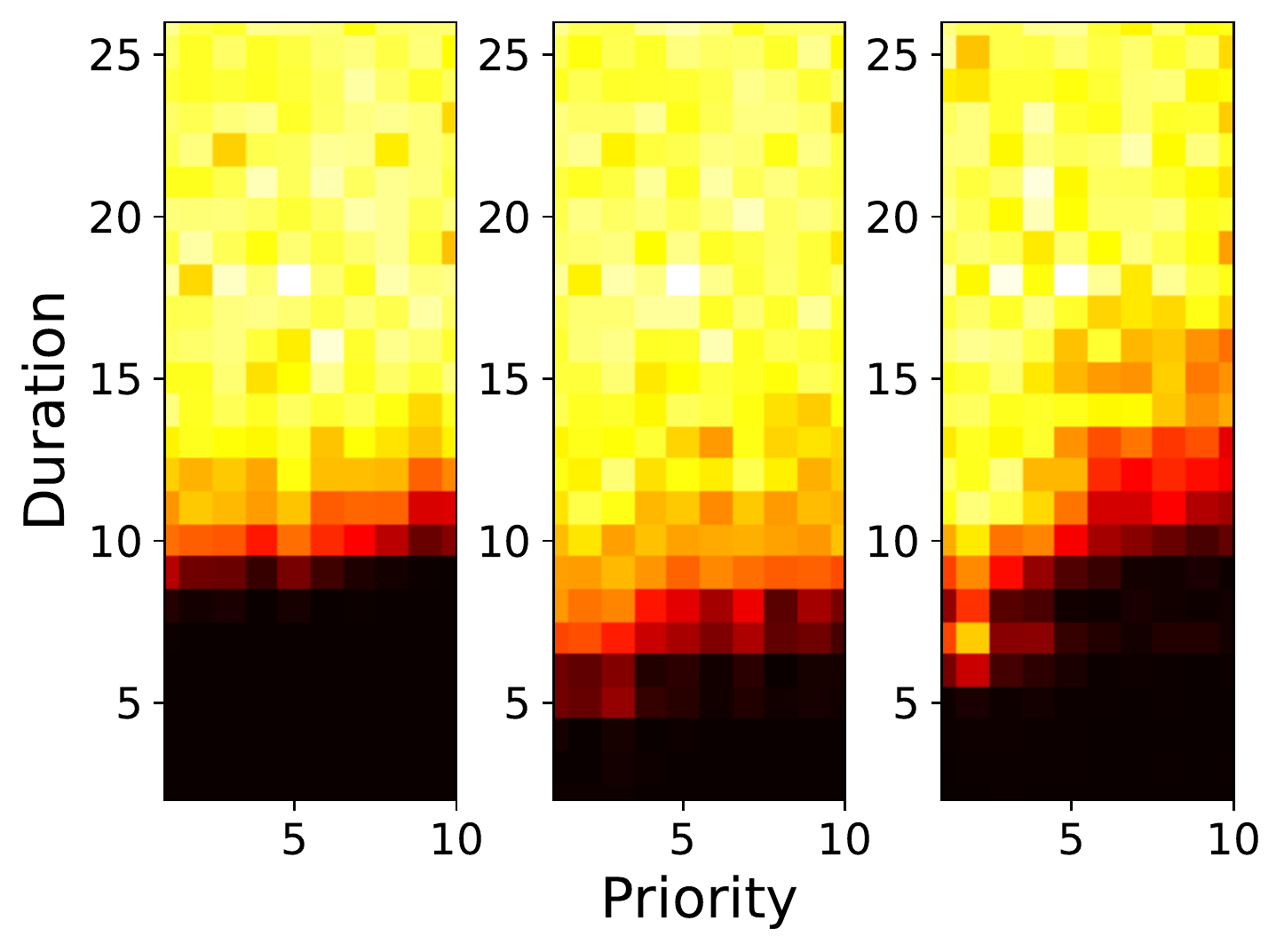}
                        \label{fig:cp_analysis_fn}
                    }
                \end{tabular}
                &
                \begin{tabular}[m]{c}
                    \subfloat{%
                        \centering
                        \includegraphics[width=0.15\linewidth]{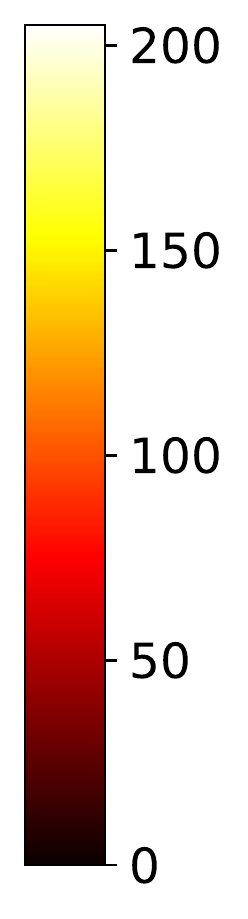}
                        \label{fig:cp_analysis_colorbar}
                    }
                \end{tabular}
            \end{tabular}
            \caption{\textit{true positives} and \textit{false negatives} with $\constantDuration = 1.8$ and $\maxDuration = 26$.}
             \label{fig:cp_analysis}
          \end{figure}

        We extend our analysis, in  Fig.~\ref{fig:cp_analysis}, to investigate which \acp{VNR} are getting accepted or rejected during evaluation.
        Fig.~\ref{fig:cp_analysis_tp} shows the number of \textit{true positives} while Fig.~\ref{fig:cp_analysis_fn} shows the number of \textit{false negatives} using different values for $\constantPriority$ during training.
        Each point represents the number of embedded \acp{VNR} with their corresponding duration and priority values.
        For the agent with $\constantPriority = 1.2$ most \textit{true positives} are short \acp{VNR} of different priorities.
        As the \ac{VNR} duration increase, they are being rejected by the admission control. VNRs with longer duration are not embedded at all, which can be seen in the dark area.  The results for \textit{false negatives}, in Fig.~\ref{fig:cp_analysis_fn} confirm this behavior.
        Mainly long \acp{VNR} are being rejected, even if they can be embedded.

        Higher values for $\constantPriority$ have two obvious impacts on the acceptance rate (Fig.~\ref{fig:cp_analysis_tp}).
        First, the slope of the red boundary increases -- the agent starts to accept longer \acp{VNR} with high priorities rather than short \acp{VNR} with low priority.
        Second, the dark area in the \textit{true postive} graphs gradually changes to red.
        That means the agent has no longer hard constraints on rejecting long \acp{VNR} and becomes more flexible toward accepting long \acp{VNR}. This, however, decreases the long-run acceptance rate: the average acceptance rate decreases to be 74\%, 48\%  and 40\% when increasing $\constantPriority$ to  1.2, 1.5 and 1.8,  respectively. Meanwhile the average priority of those embedded \acp{VNR} increases from 5.4 ($\constantPriority=1.2$) to 6.1 ($\constantPriority=1.8$), while those who were rejected even though there were available resources had on average 5.4 ($\constantPriority=1.2$) and 5.1 ($\constantPriority=1.8$) priorities.

\begin{figure}
        	\centering
        	\subfloat[Acceptance rate]{%
        		\centering
        		\includegraphics[width=0.48\linewidth]{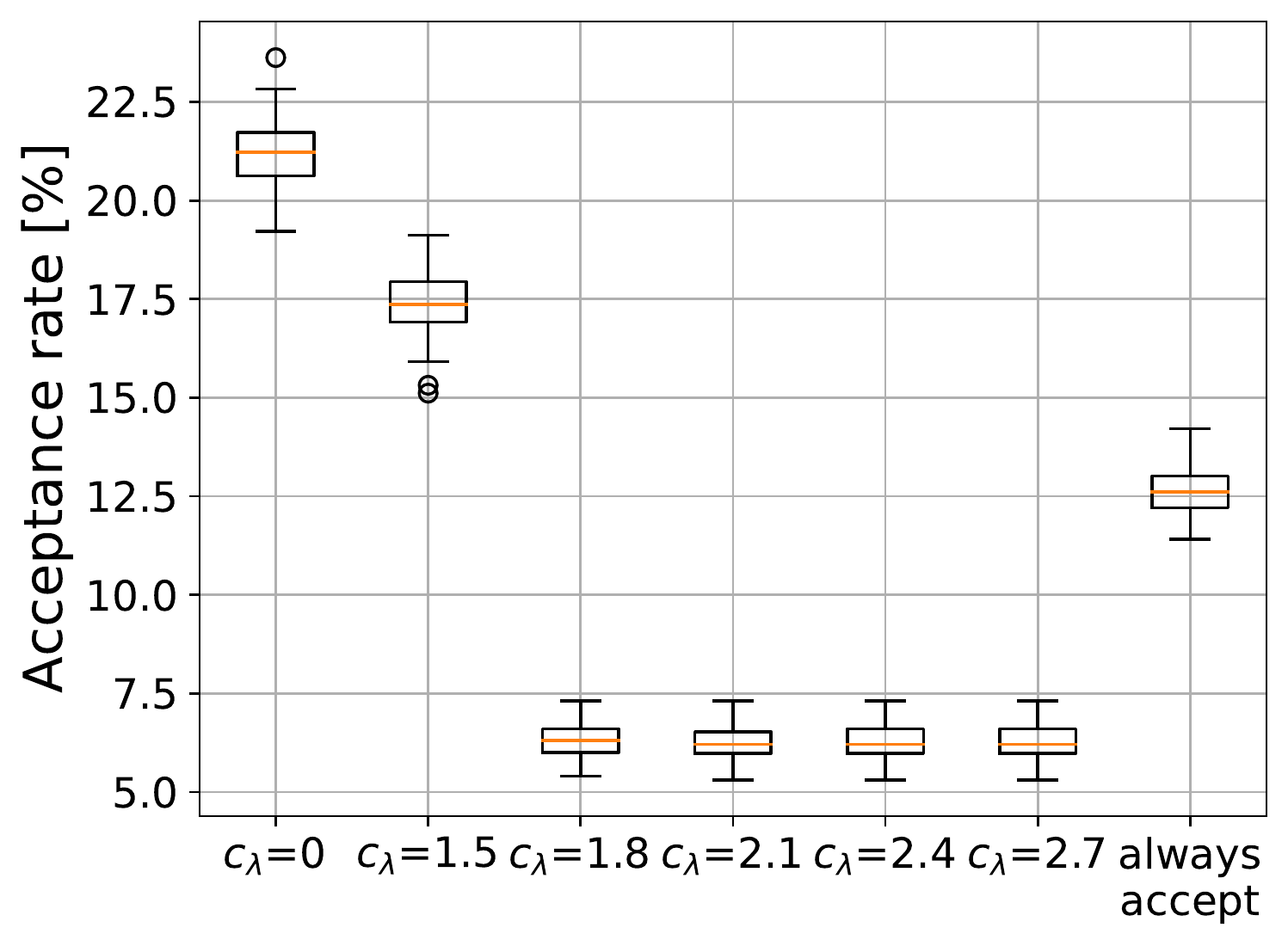}
        		\label{fig:boxplot_tp2}
        	}
        	\subfloat[False negatives]{%
        		\centering
        		\includegraphics[width=0.48\linewidth]{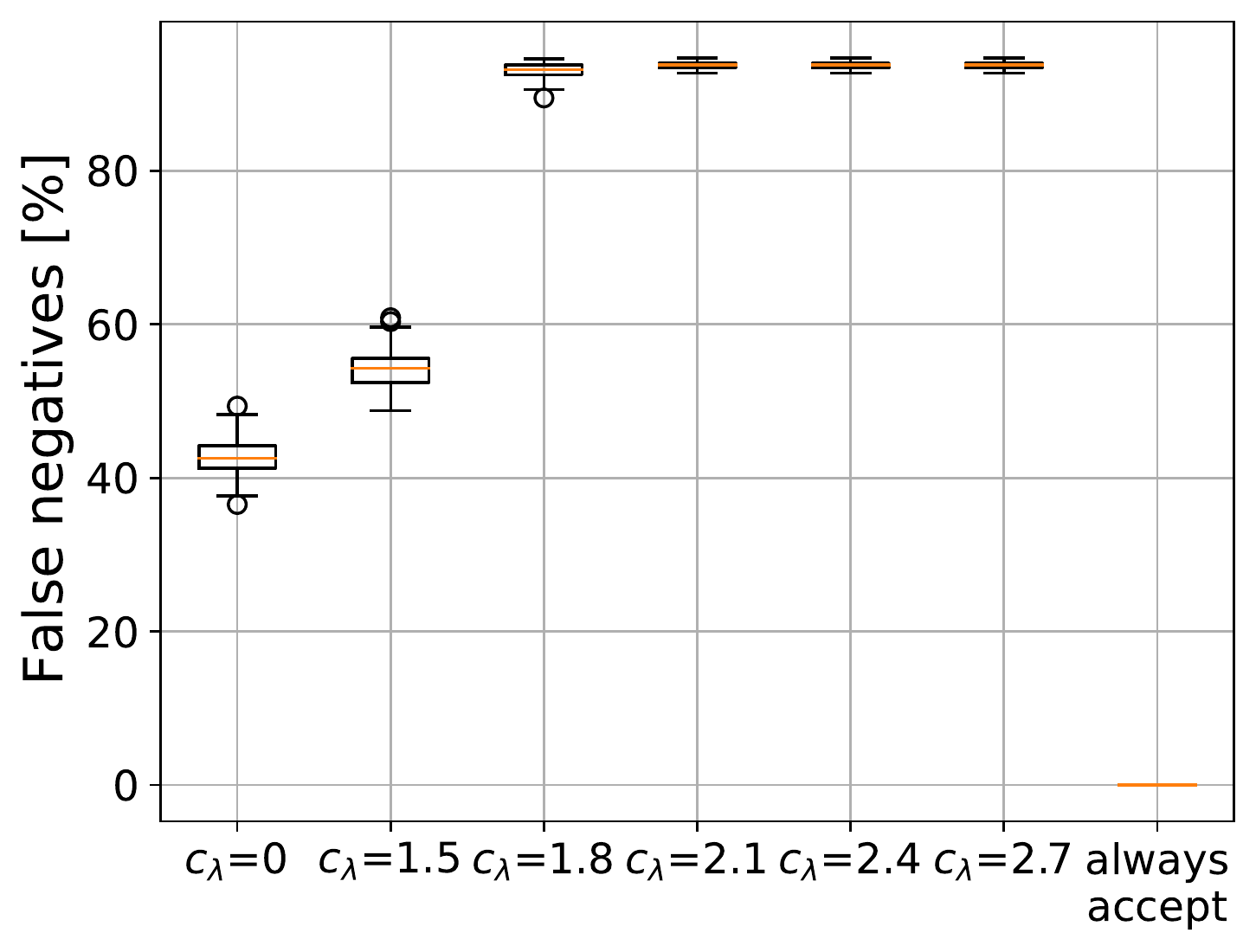}
        		\label{fig:boxplot_fn2}
        	}
        	\caption{Results of $\constantPriority$ parameter analysis with $\constantDuration = 1.8$ and $\maxDuration = 30$}
        	\label{fig:boxplots2}
        \end{figure}

        Meanwhile, $\constantPriority$ should be carefully tuned, otherwise it will lead to an undesired agent behavior. In Fig.~\ref{fig:boxplots2}, we retrain our environment for $\constantDuration=30$ and, again, with increasing $\constantPriority$. We observe that 
        agents trained with high values for $\constantPriority$ eventually start to  prioritize \textit{false negatives}   (Eq.~\eqref{eq:extra_reward}). Hence, the results could be even worse than the baseline solution: lower acceptance rate (Fig.~\ref{fig:boxplot_tp2}) and higher false positive (Fig.~\ref{fig:boxplot_fn2}).
        
        
        \subsection{Changing maximum trained duration}
        In the previous sections, agents are trained with a fixed $\maxDuration$ and evaluated for different incoming $\maxDuration$. \textit{What if the agents were trained for a shorter  $\maxDuration$? How would that impact the acceptance rate for incoming \acp{VNR} with longer, shorter, or equal to the trained $\maxDuration$?}
        
\begin{figure}
        	\centering
        	\subfloat[Acceptance rate]{%
        		\centering
        		\includegraphics[width=0.48\linewidth]{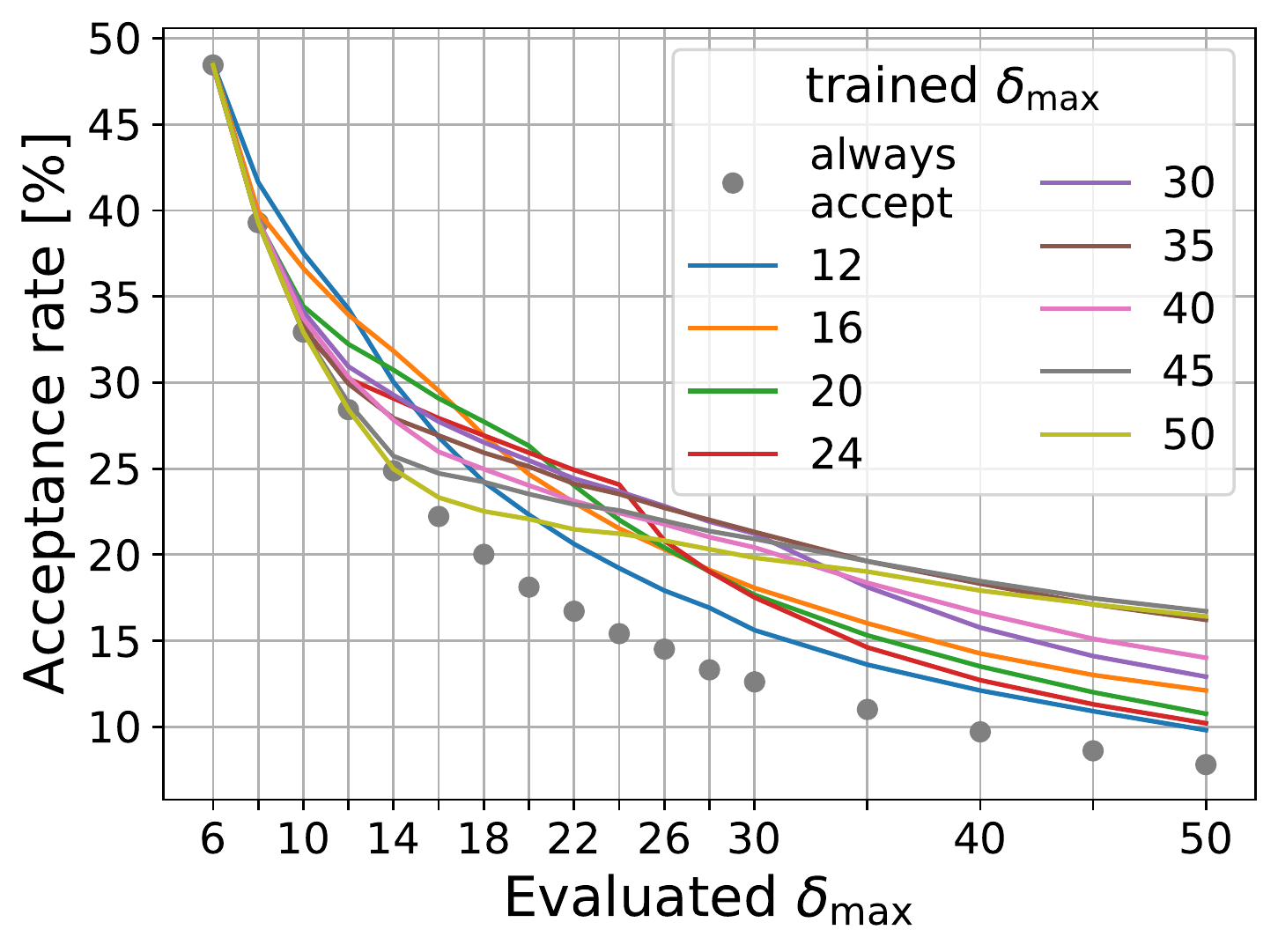}
        		\label{fig:tp_duration_control}
        	}
        	\subfloat[False negatives]{%
        		\centering
        		\includegraphics[width=0.48\linewidth]{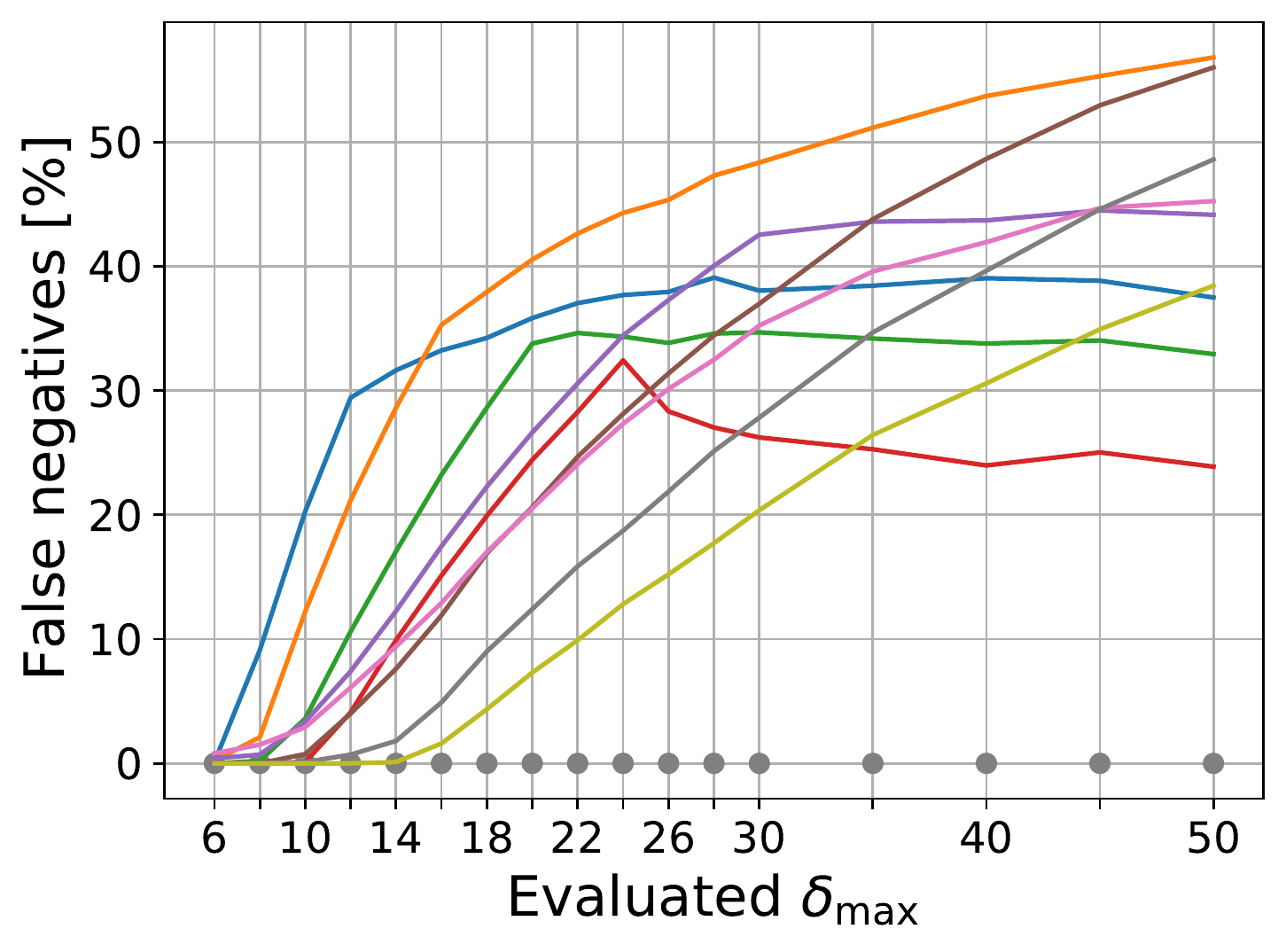}
        		\label{fig:fn_duration_control}
        	}
        	\caption{Impact of trained $\maxDuration$ on agent performance in different $\maxDuration$ enviromnets}
        	\label{fig:duration_control}
        \end{figure}

        To answer these questions, in total 9 agents are trained for $\maxDuration$ between 12 and 30 and evaluated, in addition to the baseline,  for different $\maxDuration$ (Fig.~\ref{fig:duration_control}).
        All training runs are performed with $\constantDuration = 1.8$, which is a compromise between increasing acceptance rate and limiting the number of \textit{false negatives} as described in \ref{sec:cd_values}. We set $\constantPriority=0$.
        
        In Fig.~\ref{fig:tp_duration_control}, each agent shows the highest increase in median acceptance rate
        compared to the \textit{always accept} baseline agent in the exact scenario it was trained for.
        For higher offered loads, the agents' performance drops but still stays above the baseline level.
        This behavior results from agents not having encountered any larger \ac{VNR} duration during training thus do not know how to properly deal with them during evaluation.
        
        This can also be interpreted from Table~\ref{tab:diff_env}. Agents perform mostly the best when the trained and evaluated $\maxDuration$ are the same. As the difference between the trained and evaluated $\maxDuration$ increase, the acceptance rate decrease, but it is still better than the baseline solution. Meanwhile, Fig.~\ref{fig:fn_duration_control} does not show consistent patterns for the relation between trained and evaluated $\maxDuration$ with respect to \textit{false negatives}. Hence, it depends more on the environment and should be tuned with respect to other environment parameters (e.g., distribution of arriving \ac{VNR} duration).

            \begin{table}
    \caption{Comparison between agents with different trained $\maxDuration$ in multiple $\maxDuration$ environments}
    \centering
    \resizebox{0.45\textwidth}{!}{%
    \begin{tabular}{lrrrrrrrrr}
        \hline
        eval $\maxDuration$   &   12 &   16 &   20 &   24 &   30 &   35 &   40 &   45 &   50 \\
        trained $\maxDuration$  &    &    &    &    &    &    &    &    &    \\
        \hline
        12                                             & \textbf{34.3} & 26.8 & 22.3 & 19.2 & 15.6 & 13.6 & 12.1 & 10.9 &  9.8 \\
        16                                             & 33.9 & \textbf{29.5} & 24.7 & 21.5 & 18.1 & 16.0 & 14.3 & 13.0 & 12.1 \\
        20                                             & 32.2 & 29.1 & \textbf{26.3} & 22.0 & 17.7 & 15.3 & 13.5 & 12.0 & 10.8 \\
        24                                             & 30.2 & 27.9 & 25.9 & \textbf{24.1} & 17.5 & 14.6 & 12.7 & 11.3 & 10.2 \\
        30                                             & 30.9 & 27.7 & 25.5 & 23.7 & 21.2 & 18.1 & 15.8 & 14.1 & 12.9 \\
        35                                             & 29.9 & 26.9 & 25.1 & 23.5 & \textbf{21.3} & \textbf{19.6} & 18.3 & 17.1 & 16.2 \\
        40                                             & 30.3 & 26.0 & 24.0 & 22.4 & 20.4 & 18.4 & 16.6 & 15.1 & 14.0 \\
        45                                             & 28.8 & 24.7 & 23.5 & 22.6 & 20.9 & 19.6 & \textbf{18.5} & \textbf{17.5} & \textbf{16.7} \\
        50                                             & 28.4 & 23.3 & 22.1 & 21.2 & 19.8 & 19.0 & 17.9 & 17.1 & 16.4 \\
        baseline                                       & 28.4 & 22.2 & 18.1 & 15.4 & 12.6 & 11.0 &  9.7 &  8.6 &  7.8 \\
        \hline
    \end{tabular}}
\label{tab:diff_env}
\end{table}
\section{Summary}
\label{sec:summary}
This paper uses \ac{RL} for admission control of \acp{VNR} in wireless \ac{VNE}. 
\ac{RL} agents are trained to maximize the revenue (e.g., acceptance rate or  number prioritized \ac{VNR}) by rejecting some \acp{VNR} to be able to accept more/better future \acp{VNR}.
The agent's behavior can be tuned by modifying the parameters $\constantDuration$ and $\constantPriority$.
On the one hand, low values of $\constantDuration$ lead to a behavior with little to no increase in overall acceptance rate while keeping the number of \textit{false negatives} small. High values of $\constantDuration$ yield more \textit{false negatives}  aggressively, resulting in a higher overall acceptance rate, but with a diminishing return. On the other hand,
the higher $\constantPriority$ the more likely the agent consider priority over duration.
Hence, longer \acp{VNR} with a high priority value can be accepted. This should be carefully tuned, otherwise the results can be even worse than a first fit greedy baseline. Meanwhile, training an agent with a predefined $\maxDuration$ yields a better solution than the first fit baseline, even if there is a mismatch between the trained and deployed/evaluated $\maxDuration$.

\bibliographystyle{unsrt}
\bibliography{literature}


\end{document}